\documentclass[11pt,onecolumn]{apa}
\usepackage{a4,amssymb,graphicx,eurosym,apacite,gensymb,color,listings}
\newcommand{\htr}[1]{\color{black} #1~\color{black}}

\begin{document} 

\begin{center}
{\LARGE\bf
\htr{Spatial statistics for gaze patterns in\\[0.7ex]
scene viewing: Effects of repeated viewing}\\}
\vspace{10mm}
\large
Hans A.~Trukenbrod$^{1}$ \& Simon Barthelm\'e$^{2}$  \& \\ Felix A.~Wichmann$^{3,4,5}$ \& Ralf Engbert$^{1}$\\
\vspace{10mm}
\normalsize
$^1$University of Potsdam, Germany\\
$^2$CNRS, Gipsa-lab, Grenoble INP, France\\
$^3$Eberhard Karls University of T\"ubingen, Germany\\
$^4$Bernstein Center for Computational Neuroscience T\"ubingen, Germany\\
$^5$Max Planck Institute for Intelligent Systems, T\"ubingen, Germany\\
\vspace{10mm}
(Version: \today)
\end{center}

\vspace{30mm}

\vspace{\fill}

\noindent
Address correspondence to: \\
Hans Trukenbrod \\
Department of Psychology \\
University of Potsdam\\ 
Am Neuen Palais 10\\ 
D--14469 Potsdam\\
Germany \\ 
E-mail: Hans.Trukenbrod\symbol{64}uni-potsdam.de \\
Phone: +49 331 9772703, Fax: +49 331 9772794

\newpage

\newpage

\noindent
{\bf Scene viewing is used to study attentional selection in complex but still controlled environments. One of the main observations on eye movements during scene viewing is the inhomogeneous distribution of fixation locations: While some parts of an image are fixated by almost all observers and are inspected repeatedly by the same observer, other image parts remain unfixated by observers even after long exploration intervals. Here, we apply spatial point process methods to investigate the relationship between pairs of fixations. More precisely,  we use the \emph{pair correlation function} (PCF), a powerful statistical tool, to evaluate dependencies between fixation locations along individual scanpaths. We demonstrate that aggregation of fixation locations within four degrees is stronger than expected from chance. Furthermore, the PCF reveals stronger aggregation of fixations when the same image is presented a second time. We use simulations of a dynamical model to show that a narrower spatial attentional span may explain differences in pair correlations between the first and the second inspection of the same image.
} 

\section*{Introduction}

When we move our eyes during scene viewing, fixation locations are not selected \htr{uniformly.} Instead fixations cluster in specific areas while other areas remain unfixated by observers, even after long exploration intervals. Most research on this inhomogeneity has tried to identify factors that contribute to the placement of fixations across trials, while statistical correlations within trials have mostly been ignored. Here we describe the pair correlation function (PCF), a method from spatial statistics, to investigate the relationship between fixation locations of individual scanpaths. In a first part, we \htr{provide a step-by-step tutorial how the PCF can be applied to eye movement data and demonstrate} that fixations in a scanpath are more aggregated than expected by the distribution of fixation locations of all subjects \htr{\cite{Engbert.JVis.2015}.} In a second part, we show that a long-term memory manipulation, i.e. the second inspection of an image, leads to even more aggregation. We discuss the results in the light  of simulations of the SceneWalk model \cite{Engbert.JVis.2015}, a dynamical model for the generation of saccadic sequences in scene viewing. \htr{Furthermore, we demonstrate with simulated fixation locations that the PCF can be used to test specific hypotheses and that seemingly similar distributions of fixation locations, may lead to very different PCFs.} 

\subsection*{Eye movements during scene perception.}
\htr{Fixation locations during scene perception are influenced by bottom-up and top-down as well as low-level and high-level factors \cite<see>[for an extensive discussion]{Schutt.arXiv.2018}. Bottom-up factors refer to parts of the image that attract gaze independent of the internal state of an observer and might differ in complexity. Simple low-level features are extracted early in the visual hierarchy while high-level features are complex shapes and extracted late in the visual hierarchy. Examples of bottom-up, low-level features that predict fixation locations are luminance contrast and edge density \cite{Mannan.Perception.1997,Reinagel.NetworkCompNeural.1999,Tatler.VisionRes.2005}. The strength of the relationship for different image features depends on the type of image viewed \cite{Parkhurst.VisionRes.2002}. Examples of bottom-up, high-level features that predict fixation locations are objects \cite<e.g., faces, persons, cars;>{Einhauser.JVis.2008,Cerf.AdvNeuralInfoProcSyst.2007,Judd.CompVision.2009}. This interpretation is supported by the existence of a preferred viewing location close to the object center in scene viewing \cite{Nuthmann.JVis.2010}.

The influence of bottom-up factors has led to the development of computational saliency models (\citeNP{Itti.NatRevNeurosci.2001}; cf., \citeNP{Koch.HumNeurobiol.1985}) and a variety of models has been put forward over the years \cite<e.g.,>{Bruce.JVis.2009,Kienzle.JVis.2009,Kummerer.CoRR.2016,Parkhurst.VisionRes.2002,Vig.CVPR.2014,Zhang.JVis.2008}. In particular since the rise of sophisticated machine-learning algorithms, these models perform well in predicting fixation locations when evaluated with a data set obtained under unconstrained (``free'') viewing \cite<>{Bylinskii.SaliencyBenchmark}.

Top-down factors refer to cognitive influences on fixation locations and depend on the internal state of an observer (e.g., aims, memory load, knowledge). As for bottom-up factors, top-down factors differ in respect to the complexity of features. Examples of top-down, low-level factors can be found in visual search, where observers search for a specific color or a line of a specific orientation. Examples of top-down, high-level factors are task instructions where observers need to judge the age of people or their wealth. In his seminal work, \citeA{Yarbus.Book.1967} reported anecdotal evidence for top-down control. Scanpaths were influenced by the instruction given before viewing an image \cite<see also>{Castelhano.JVis.2009}. These top-down effects strengthen when observers are engaged in natural tasks like preparation of a sandwich or during tea-making \cite{Hayhoe.TrendsCognSci.2005,Land.VisionRes.2001}. In natural tasks, fixations generally support the smooth execution of a task and occur on objects just-in-time \cite{Ballard.BehavBrainSci.1997} or as look-ahead fixations to inspect objects needed later during the task \cite{Pelz.VisionRes.2001}. Thus, the eyes do not necessarily fixate the most salient  location (bottom-up) but are rather directed towards informative locations that are important for task execution. Another source of top-down control comes from memory representations due to reinspection of previously seen images \cite{Kaspar.PLoSOne.2011,Kaspar.JVis.2011} and due to the acquired scene or world knowledge \cite{Torralba.PsycholRev.2006}. Incorporating such knowledge by contextual priors improves the predictions of saliency models \cite{Torralba.PsycholRev.2006,Judd.CompVision.2009}.}

\htr{In addition, fixation locations depend on systematic tendencies that are common to eye movements in general \cite{Tatler.JEMR.2008}. These systematic tendencies lead to spatial and directional biases in the selection of fixation locations.} A well-known example is the central fixation bias during scene viewing \cite{Tatler.JVis.2007}. On average, participants prefer to fixate near the center of an image rather than towards the periphery. \htr{The central fixation bias is strongest in the beginning of a trial and reaches an asymptotic level after a few fixations \cite{Rothkegel.JVis.2017}. Other systematic tendencies include the preference to generate positively skewed, long-tailed distributions of saccade amplitudes or the preference to execute saccades in the cardinal directions during scene viewing \cite{Tatler.JEMR.2008}.} The later effect is shaped by image features and varies systematically with the perceived horizon. Tilting an image results in an equally tilted distribution of saccade directions \cite{Foulsham.VisionRes.2008,Foulsham.VisionRes.2010}. \htr{In addition, observers tend to make saccades in the same direction as the preceding saccade (cf., saccadic momentum) and a large number of fixations bring the eyes back to the the last or penultimate fixation location \cite<cf., facilitation of return;>{Smith.VisCogn.2009,Wilming.PLoSCompBiol.2013}. 
Finally, inhibition of return is believed to facilitate exploration during visual search \cite{Klein.PsycholSci.1999} and has been suggested as a mechanism to drive attention during scene perception \cite{Itti.NatRevNeurosci.2001}. Inhibition of return during scene perception seems to primarily prolong fixation durations before return saccades to previously fixated locations \cite{Hooge.VisionRes.2005,Smith.VisCogn.2009} and seems to shape spatial dynamics of scanpaths in the long run \cite{Rothkegel.VisionRes.2016}. Adding systematic tendencies to saliency  models further improves their predictions \cite{LeMeur.VisionRes.2015}.
}



\subsection*{Eye movements and long-term memory: Repeated presentation}
Humans have a remarkable capacity to store images in long-term memory \cite{Brady.ProcNatlAcadSci.2008,Konkle.PsycholSci.2010,Standing.PsychonSci.1970}. These representations are not limited to the gist of a scene but include abstract representations of objects, in particular of previously fixated objects \cite{Hollingworth.PsychonBullRev.2001,Hollingworth.JExpPsycholHuman.2002}. By viewing the same image multiple times scene memory accumulates \cite{Melcher.Nature.2001,Melcher.VisionRes.2001}. Due to this scene-specific memories repeated presentations can be used to investigate effects of long-term memory on eye movements.

In two studies, \citeauthor{Kaspar.PLoSOne.2011,Kaspar.JVis.2011} studied the interaction of bottom-up and top-down influences and long-term memories by presenting images five times. In general, fixation locations tend to be similar across different inspections of the same image by the same participant \cite{Kaspar.JVis.2011}. Similarity was strongest for successive presentations and decreased with increasing distance between presentations. The correlation between fixation locations and low-level features, however, remained rather constant. In contrast, the number of fixated regions decreased after multiple presentations of images, as did the average number of fixations. Furthermore, saccade amplitudes were largest during the first presentation and decreased on subsequent presentations. 

In a second study \citeA{Kaspar.PLoSOne.2011} explored the effects of repeated presentations on top-down influences on saccade selection. Motivation (as measured via the reported interestingness of the viewed images) and a personality trait of participants (action orientation) influenced repeated viewing of images. In addition, fixation durations and variability between participants' fixation locations increased, whereas saccade frequency, saccade length, and entropy of fixation locations decreased. The authors concluded that the locus of attention became increasingly local with repeated presentations of images. Thus, participants scrutinized individual regions during later presentations. This interpretation was supported by participants' self-reports and was augmented in participants who found the images more interesting. 

\subsection*{Research Questions}
Much progress has been made to understand where observers fixate in an image. This research primarily focused on fixation locations across observers while neglecting the fixation history during a trial of a single observer. Fixation locations, however, exhibit strong spatial correlations during a trial and are not independent of one another \cite{Engbert.JVis.2015} and adding mechanisms that generate more realistic scanpaths improves performance of saliency models \cite<>{LeMeur.VisionRes.2015}. \citeA{Barthelme.JVis.2013} introduced spatial point processes as a theoretical framework for the study of gaze patterns, and demonstrated how this helps to turn qualitative into quantitative questions. Here, we present a method from spatial statistics, i.e., the \emph{pair correlation function} (PCF), to estimate spatial correlations between fixation locations during a trial in the presence of spatial inhomogeneity \cite{Engbert.JVis.2015}. Before applying the PCF to eye movement data during the first and second inspection of an image, we briefly describe the theoretical details of the PCF. We demonstrate (i) that the PCF provides rigorous statistical evidence for aggregation of fixation locations in single trials, \htr{(ii) that this effect is not well explained by the tendency of participants to generate short saccade amplitudes,} (iii) that the PCF is differentially affected by a memory manipulation (first vs.~second inspection of an image), and (iv) that these differences can be explained by modulations of the attentional span within our SceneWalk model \cite{Engbert.JVis.2015}.

\section*{Pair Correlation Function}
We refer the reader to \citeA{Diggle.Book.2013} for an introduction to the statistical analysis of point patterns and to  \citeA{Law.JEcol.2009} for a detailed description of the PCF and the application to point patterns in plant ecology.

\subsection*{Analyzing spatial point patterns: Pair correlation function}
The density estimation of a point pattern, i.e., the probability of observing a point at a given location, is a first-order statistic for a spatial point process, which, therefore, plays the role of the mean value in classical statistics. In the upcoming sections we denote this first-order statistic as the intensity $\lambda(x)$. In the case of eye movements the intensity represents the local average spatial density of fixations at a location $x$. Point patterns that are generated by a homogeneous point process are uniformly distributed and the underlying intensity $\lambda(x) = \lambda$ is constant for all $x$. For inhomogeneous point processes, where the 2D density of fixation locations is non-uniformly distributed, the intensity $\lambda(x)$ is estimated for each location $x$ separately, i.e., 

\begin{equation}
  \lambda(x)=\lim\limits_{|dx| \rightarrow 0}{\left\lbrace\frac{E[N(dx)]}{|dx|}\right\rbrace}.
  \label{eqIntensity}
\end{equation}
Here, the density is given by the expected number of fixations \htr{$E[N]$} falling into a disc of infinitesimal size $|dx|$. 

While first-order statistics are concerned with locations of single points (while ignoring spatial correlations between points), second-order statistics describe the relation between pairs of points. A crucial second-order statistic for the computation of PCFs is the pair density $\rho(r)$. The pair density \htr{(or second-order intensity function) describes the probability of simultaneously observing points generated by a point process} in two disjoint discs  with centers $x$ and $y$ of infinitesimal size $dx$ and $dy$,

\begin{equation}
  \rho(x,y) =\lim\limits_{|dx|,|dy| \rightarrow 0}{\left\lbrace\frac{E[N(dx)N(dy)]}{|dx||dy|}\right\rbrace}.
  \label{eqPairDensity}
\end{equation}
For a stationary, isotropic point process the pair density $\rho(x,y) = \rho(r)$ depends on the distances of pairs of points only, where $r$ corresponds to the distance between pairs of points $\|x-y\|$ \cite{Diggle.Book.2013}. Mathematically we estimate the pair density $\hat{\rho}(r)$ at distance $r$ by

\begin{equation}
  \hat{\rho}(r) = \sum_{x,y\in W}^{\neq} \frac{k(\|x-y\|-r)}{2\pi rA_{\|x-y\|}} 
  \label{eqPairDensity}
\end{equation}
where $k$ represents an Epanechnikov kernel\footnote{The Epanechikov kernel $\epsilon(x) = \frac{3}{4w} (1- \frac{x^2}{w^2})_+$ with $(x)=max(0,x)$ is a quadratic function that is truncated to the interval [-w,w].} \htr{\cite{Baddeley.Book.2015}} and $A_{\|x-y\|}$ is an edge-correction factor to counter the loss of pairs of points near the boundary of the inspection window, which is particularly important at large distances $\|x-y\|$. \htr{As edge correction we chose to use the translation correction, that weights each pair of points $(x_i,x_j$) with the reciprocal of the fraction of the window area, in which a first point $x_i$ could be placed, so that both points $x_i,x_j$ would be observable \cite{Baddeley.Book.2015} } 

Since the pair density depends on the number of points, the PCF is computed as a normalized version of the pair density (i.e., the PCF is given by the intensity-weighted pair density). In the case of a homogeneous PCF the pair density is weighted by a constant at all distances,

\begin{equation}
  g_{hom}(r)=\frac{1}{{\lambda}^2}{\rho}(r).
  \label{eqHomPcf}
\end{equation}

However, fixation locations are not distributed homogeneously and show aggregation due to bottom-up, top-down, and systematic oculomotor factors. Fortunately, the PCF can be computed for inhomogeneous processes by weighting the pair density with the intensities $\lambda(x_i)$ of an inhomogeneous point process at each fixation location,

\begin{equation}
  {g}_{inhom}(r) = \sum_{x,y\in W}^{\neq}
  \frac{1}{{\lambda}(x){\lambda}(y)}{\rho}(r)
  \label{eqInhomPcf}
\end{equation}

The resulting PCFs will be non-negative, $g(r) \ge 0$, at all distances $r$. Values of the PCF close to one, $g(r)\approx 1$, indicate that pairs of points at distance $r$ are independent. Points at distance $r$ occur solely due to the underlying intensity $\lambda(x)$. For larger values, i.e., $g(r)>1$, point patterns are more abundant at distance $r$ than expected by the intensity $\lambda(x)$. Thus, pairs of points at distance $r$ interact and observing a point $x$ increases the probability of observing a point $y$ at distance $r$. \htr{The probability of observing point $y$ is higher} than predicted by the local intensity $\lambda(y)$. Conversely, smaller values, $g(r) < 1$, reveal that points are less abundant than the spatial average at distance $r$. Observing a point reduces the probability of observing a second point at distance $r$.

\begin{figure*}
\begin{center}
\includegraphics[width=.9\linewidth]{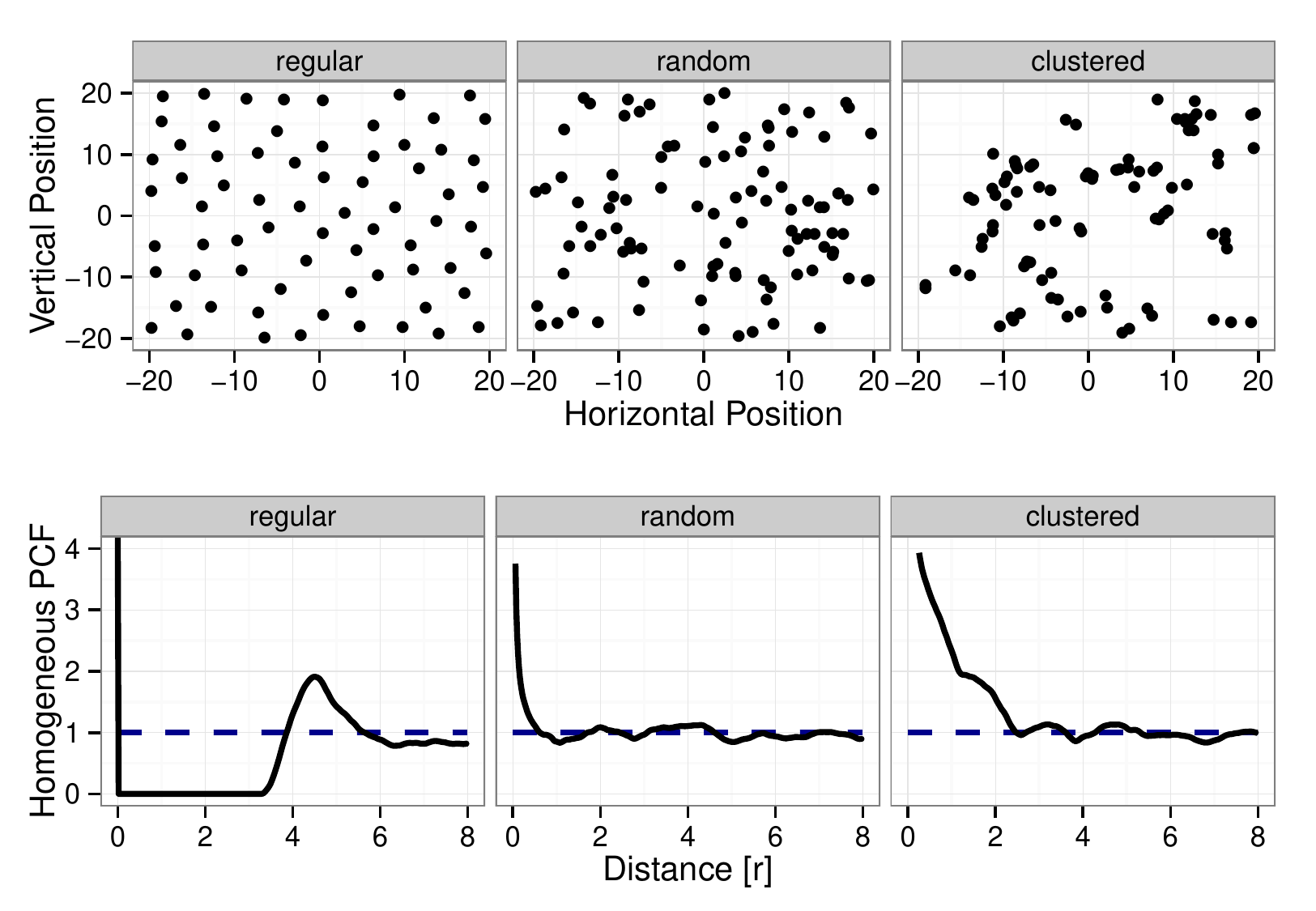}
\caption{Examples of point patterns (upper panels): regular, random, and aggregated (clustered). Corresponding pair correlation functions (lower panels).}
\label{pcfExamples}
\end{center}
\end{figure*}

Figure~\ref{pcfExamples} shows the PCF of three different point patterns. In all examples we computed the PCF assuming a homogeneous point process with constant intensity $\lambda(x) = \lambda$ (cf.,~Eq.~\ref{eqHomPcf}) since deviations from uniformity are easier to interpret visually. The same interpretation, however, can be applied to inhomogeneous PCFs. The first example shows a regular point pattern (left column). Visual inspection of the points indicate a grid-like arrangement. The distance between neighboring points is relatively constant. The resulting PCF (bottom row) summarizes this behavior. At short distances, $r<4$, the PCF reveals a strong inhibitory effect, ${g}(r) \approx 0$. The existence of a point impedes the occurrence of other points within this radius. At medium distances, $4<r<6 $, the PCF reveals aggregation of points, ${g}(r) > 1$. Observing a point boosts the occurrence of points at this distance. Hence, the grid-like appearance. At larger distances, $r>6$, the PCF lends support to the hypothesis of independence of points, since ${g}(r) \approx 1$. We observe no long-range interaction of pairs of points and the distribution of distances can be explained by the density distribution, $\lambda(x)$.

The second example (Fig.~\ref{pcfExamples}, central panels) shows \htr{the realization of a point pattern with complete spatial randomness (CSR)}. Points are distributed uniformly. The resulting PCF reveals the independence of points at all distances, ${g}(r) \approx 1$. Note, the aggregation at short distances $r$ is an artifact generated by the estimation process. Finally, the third example illustrates an aggregated point process (right panels). The PCF at short distances, $r<2$, reveals aggregation, ${g}(r) > 1$, while the PCF at longer distances reveals independence,  ${g}(r) \approx 1$. Thus, observing a point increases the likelihood of observing other points in close proximity. The occurrence of distant points can be explained by the uniform distribution.

Finally, deviations from \htr{complete spatial randomness}, i.e., $g(r) \neq 1$, can be summed up and serve as a useful summary statistic of the overall behavior:

\begin{equation}
\chi=  \int_0^\infty (g(r)-1)^2 dr.
  \label{eqTestStat}
\end{equation}
In practical applications, the deviation from \htr{complete spatial randomness} $\chi$ is computed over a finite interval, e.g., $\int_a^b(g(r)-1)^2dr$ with $a<b$.

\subsection*{Application of the PCF to fixation locations}
In this section we demonstrate how to compute the \emph{pair correlation function} (PCFs) for eye movement data\htr{ in three steps.} The PCF \htr{reveals whether the distribution of fixation locations during a single scanpath can be explained by the overall inhomogeneity observed across all observers or whether fixation locations of a single scanpath contain additional spatial correlations}. All analyses and graphs reported have been implemented in \emph{R} using the  \emph{spatstat} \cite{Baddeley.JStatSoft.2005,Baddeley.Book.2015} and \emph{ggplot2} packages \cite{Wickham.Book.2009}. We provide R-code at \url{http://www.rpubs.com/hans/PCF}.

\subsubsection{1. Simulate inhomogeneous and homogeneous control processes}
\htr{To evaluate our PCF computations, we simulate two control point processes, namely a homogeneous and an inhomogeneous point process. Points (fixation locations) are sampled independently from each other in both control processes and due to the independence of points, we do not expect to observe any correlations between points at distance $r$. Any observed correlations would be spurious and depend on the data structure (e.g., length of fixation sequences) or a wrong parameterization of the method. Hence, both control processes ensure that correlations in the PCF arise from the empirical data and not by the method itself. In addition, the inhomogeneous point process is used in the second step to estimate an optimal bandwidth for the intensity estimation of the PCF.}

\begin{figure*}
\begin{center}
\includegraphics[width=.9\linewidth]{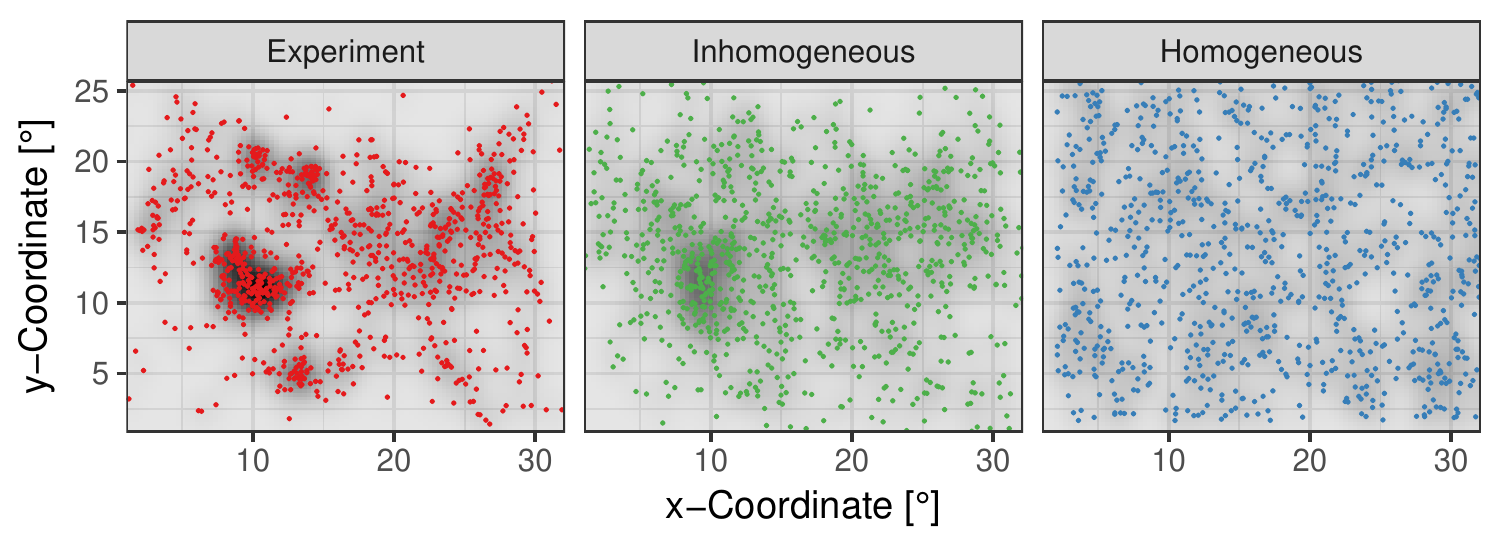}
\caption{Distribution of fixation locations. Dots represent fixation locations, estimated intensities of each point pattern are illustrated by gray shading. Experimental fixation locations (red, left panel), locations generated by a inhomogeneous point process (green, central panel), and a homogeneous point process (blue, right panel).}
\label{pcfDensities}
\end{center}
\end{figure*}

Figure~\ref{pcfDensities} (left panel) shows fixations on an image from participants viewing the same scene for 10 seconds (see Methods for details). As expected fixation locations are not uniformly distributed and indicate inhomogeneity. The estimated intensity $\hat{\lambda}_s(x)$ of all fixation locations is depicted by gray shading where darker areas represent higher intensities. \htr{We used Scott's rule of thumb to compute the smoothing bandwidth for the intensity estimation (R-function: {\tt bw.scott}) from the \emph{spatstat} package \cite{Baddeley.JStatSoft.2005} and} estimated an optimal bandwidth for each image. The estimated intensity $\hat{\lambda}_s(x)$ is used to simulate an inhomogeneous control point process. The inhomogeneous point process (central panel) samples points proportionally to the intensity $\hat{\lambda}_s(x)$. Hence, the resemblance of the experimental and simulated distributions. The homogeneous point process (right panel) samples from a uniform distribution across the entire image area, therefore the subsequently estimated intensity (gray shading) is approximately constant, $\hat{\lambda}_s(x) \approx \lambda$. For every empirical scanpath we simulated one scanpath of equal length (same number of fixations) for the inhomogeneous point process and for the homogeneous point process obtained from Monte-Carlo simulations.

\begin{figure*}
\begin{center}
\includegraphics[width=.9\linewidth]{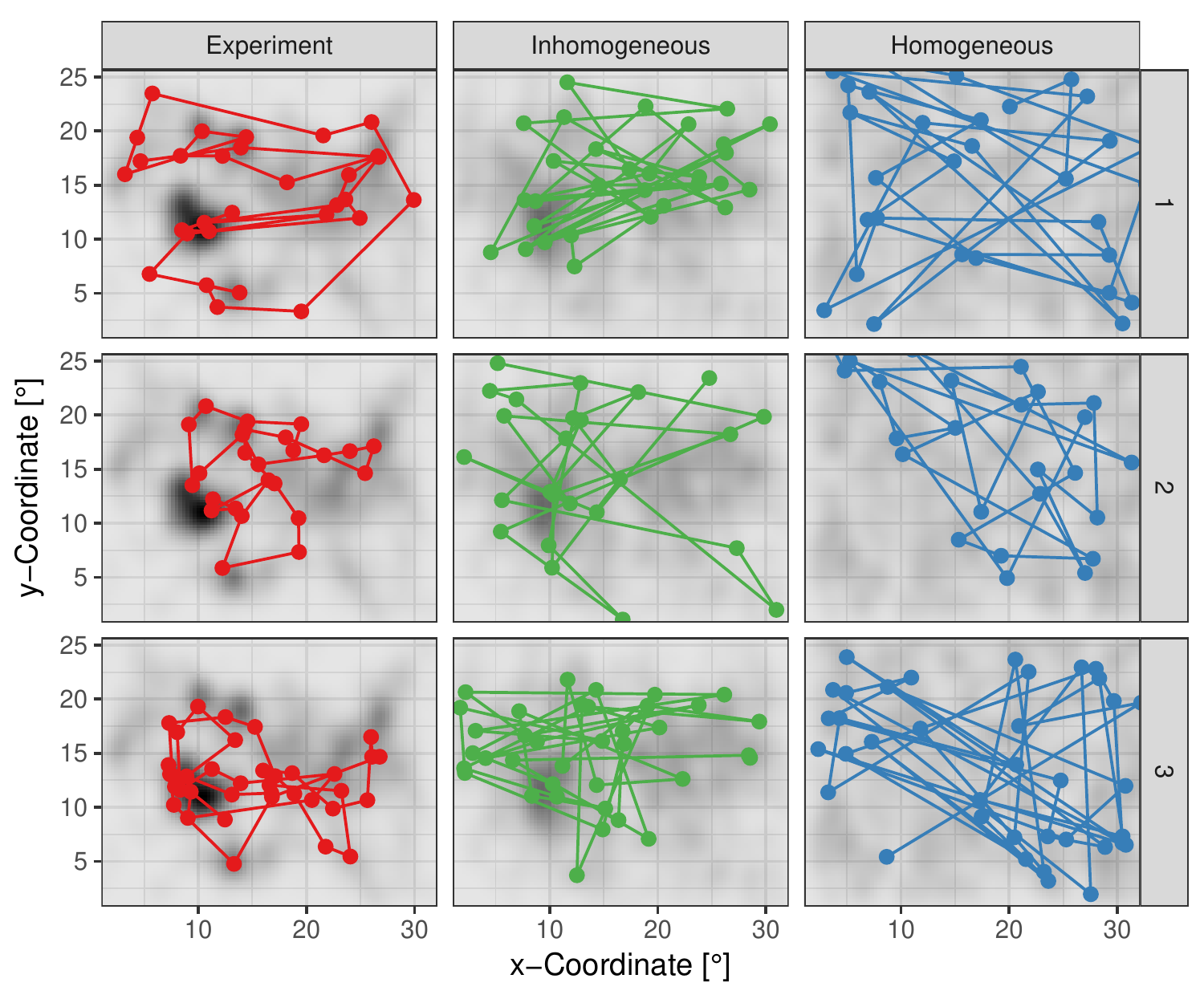}
\caption{Three representative scanpaths during viewing of an image. Each row represents the scanpath during a single trial of the empirical data (red, left panel), the simulated inhomogeneous point process (green, central panel) and the simulated homogeneous point process (blue, right panel).}
\label{pcfTrials}
\end{center}
\end{figure*}

Examples of the empirical and simulated scanpaths are visualized in Figure~\ref{pcfTrials}. Each row shows matching scanpaths of the empirical, inhomogeneous and homogeneous point process. The estimated overall intensity of each point process on an image is displayed via gray shading. Fixations are likely to be located in areas of high average intensity $\hat{\lambda}(x)$. However, each sequence consists of a unique set of points where some scanpaths explore otherwise ignored locations or ``missed'' locations of high intensity (Fig.~\ref{pcfTrials}). Overall, scanpaths of the inhomogeneous and homogeneous point processes reveal less systematic exploration behavior than the empirical data. Hence, saccade amplitudes increase considerably.

\subsubsection{2. Choose optimal bandwidth for intensity estimation of PCF} 
\begin{figure*}
\begin{center}
\includegraphics[width=.9\linewidth]{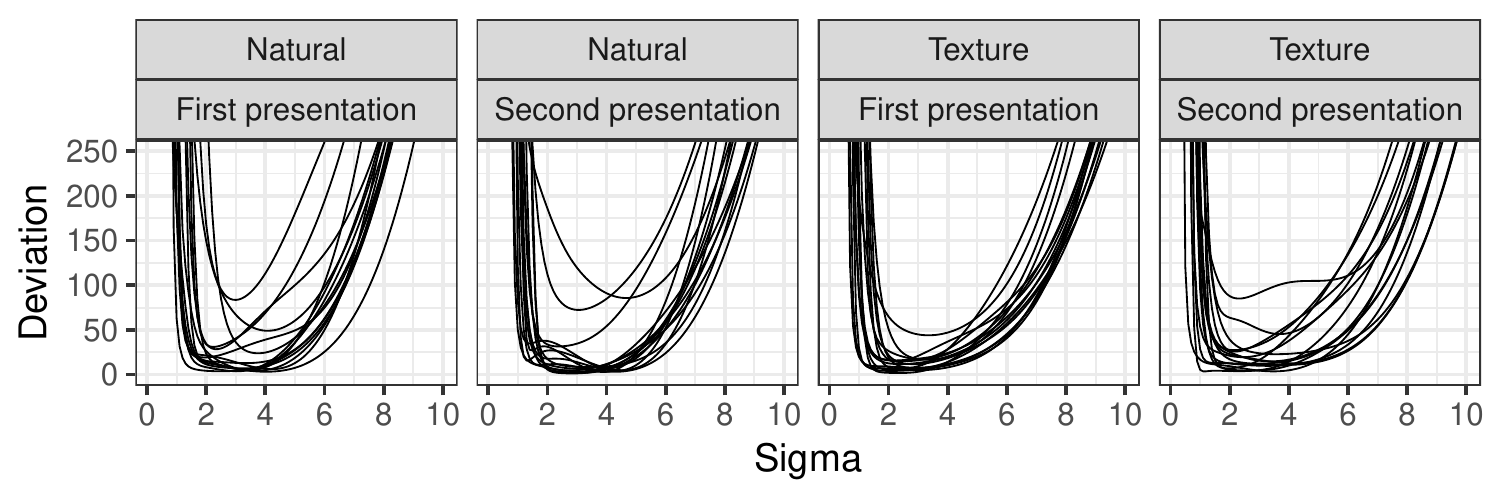}
\caption{Optimal bandwidth for intensity estimation of PCF. The deviation from complete spatial randomness (cf., Eq. \ref{eqTestStat}) is evaluated for bandwidths between 0.1\degree~and 10\degree~in steps of 0.1\degree~for each image.}
\label{pcfOptimalSigma}
\end{center}
\end{figure*}

Next we need to choose an optimal bandwidth for estimation of the intensity $\hat{\lambda}(x)$ used to calculate the inhomogeneous PCF $\hat{g}(r)$ (see Eq. \ref{eqInhomPcf}). \htr{Note, this is different from the bandwidth estimated in Step 1 to simulate the inhomogeneous point process.} Since fixation locations in scanpaths of both control point processes are sampled independent from the preceding fixation history, average PCFs of both point processes are expected to reveal no spatial correlations, i.e. we expect $\hat{g}(r)\approx 1$ at all distances $r$. We computed the deviation from \htr{complete spatial randomness} (Eq. \ref{eqTestStat}) for PCFs computed with different bandwidths. We varied bandwidths from $0\degree$ and $10\degree$ in steps of $0.1\degree$ and computed the deviation from \htr{complete spatial randomness} for  each scanpath. The average deviation at each bandwidth is plotted in Figure~\ref{pcfOptimalSigma}. Lines represent individual images. For all images, the deviation increases for small bandwidths and large bandwidths with an optimal bandwidth between $1.5\degree$~and $5\degree$.  The bandwidth yielding the smallest deviation was chosen for the intensity estimation of the PCF.

\subsubsection{3. Compute PCF for each trial}
\begin{figure*}
\begin{center}
\includegraphics[width=.9\linewidth]{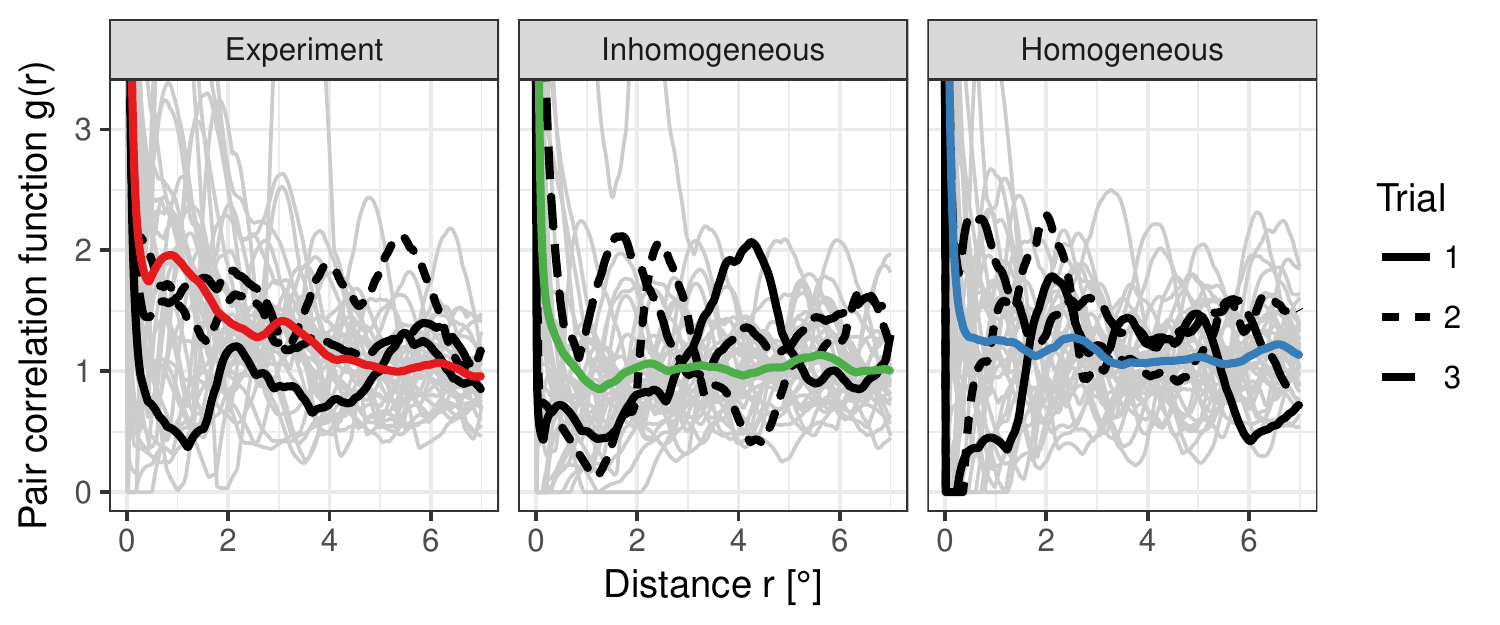}
\caption{\htr{Pair correlation functions of individual scanpaths (gray lines) and the average PCF on an image for the experimental data (red, left panel), inhomogeneous point process (green, central panel) and homogeneous point process (blue, right, panel). PCFs of trials from Figure~\ref{pcfTrials} are displayed in black.}}
\label{pcfEstimated}
\end{center}
\end{figure*}

In the last step we compute the PCF of our empirical data. For estimation, we use the intensity $\hat{\lambda}(x)$ that resulted in the smallest deviation from \htr{complete spatial randomness} of the PCF  $\hat{g}(r)$ for the inhomogeneous point process (see previous step). PCFs of individual scanpaths on an image are displayed in Figure~\ref{pcfEstimated} (gray lines). The three example scanpaths from Figure~\ref{pcfTrials} are plotted in black. PCFs vary strongly between individual trials for all point processes. The average empirical PCF across all scanpaths on an image (red line) deviates from \htr{complete spatial randomness} $\hat{g}(r) \neq 1$ for distances smaller than $4\degree$. At distances beyond 4\degree~the average PCF  suggests independence of points, i.e., $\hat{g}(r)\approx 1$. Thus, fixations co-occur in close proximity during individual trials. Conversely, areas further away are fixated as predicted by chance, i.e., the overall inhomogeneity $\hat{\lambda}(x)$ observed across all participants. Inspection of the control point processes demonstrates the absence of spatial correlations. The average PCF of the inhomogeneous and homogeneous point process are constant with  $\hat{g}(r) \approx 1$. The artifact of the estimation procedure  at short distances is present in all estimates.

\begin{figure*}
\begin{center}
\includegraphics[width=\linewidth]{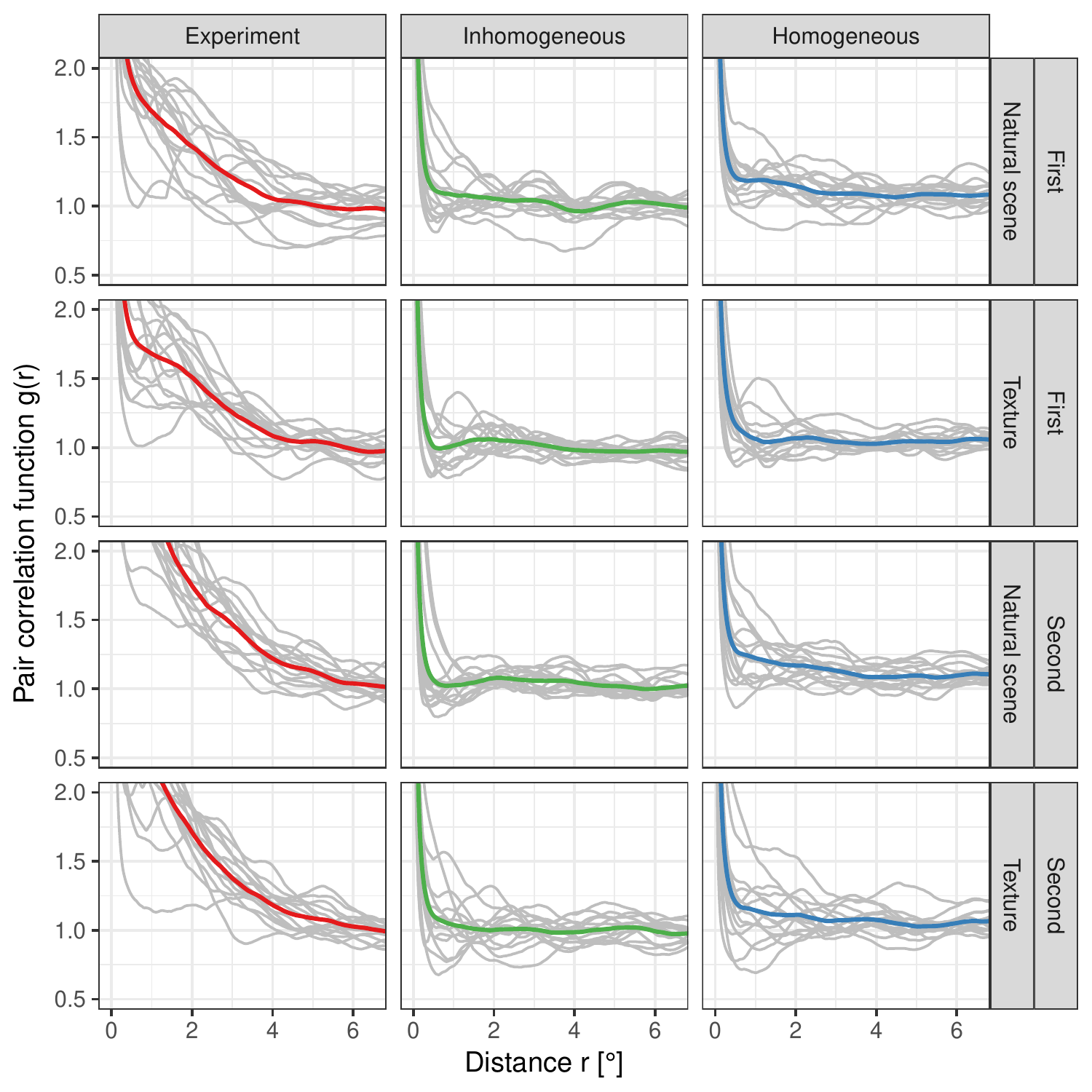}
\caption{Pair correlation functions of individual images. PCFs were estimated for each condition separately (natural vs. texture images;  first vs. second inspection). Estimated PCFs of the empirical fixation locations (red lines), simulated locations generated by a inhomogeneous point process (green lines), and an homogeneous point process (blue lines).}
\label{pcfEstimatedImage}
\end{center}
\end{figure*}

The same procedure can be repeated for each image. Figure~\ref{pcfEstimatedImage} shows PCFs of each image (gray lines) as well as the average across all images (colored lines). While inhomogeneous and homogeneous point processes reveal no spatial correlations, empirical PCFs show spatial aggregation at short distances, $r < 4\degree$ in all conditions. For a more detailed discussion see the Results section.

\section*{Methods}
\htr{For our experiment we had participants view two types of images twice. The repeated presentation of images can be understood as a form of visual long-term memory manipulation. From previous work it can be expected that this leads to similar fixation densities but shortens saccade amplitudes during the second inspection \cite{Kaspar.PLoSOne.2011,Kaspar.JVis.2011}. The resulting point patterns are similar but differ slightly in the overall inhomogeneity, which makes a direct comparison of the eye movement behavior difficult. The same problem is true for the comparison of different image types. The PCF takes differences in the underlying inhomogeneity into account and allows a direct comparison of the spatial correlations under different viewing conditions (first vs. second presentation) and for different image types (natural vs. texture images).}

\subsection*{Participants}
We recorded eye movements of 35 participants (15 male, 20 female) aged 17--36 years (mean: 24.0). Participants received study credits or 8\euro~for participation and were recruited at the University of Postdam and from a local school (32 students,
3 pupils). All participants had normal or corrected-to-normal vision as assessed by the Freiburg Vision Test \cite<FrACT;>{Bach.OptomVisSci.1996}.\footnote{\url{http://michaelbach.de/fract/}} The study conformed to the Declaration of Helsinki as well as national ethics guidelines. We obtained written informed consent from all participants.

\subsection*{Apparatus} 
Stimuli were presented on a 20\verb+"+ CRT monitor (Mitsubishi Diamond Pro 2070; refresh rate 120~Hz; resolution: 1280$\times$1024 pixels). Eye movements were recorded binocularly using the video-based Eyelink 1000 system (SR Research, Osgoode, ON, Canada) with a sampling rate of 1000 Hz. In order to reduce head movements we asked participants to position their head on a chin-rest in front of the computer screen (viewing distance: 70~cm). Stimulus presentation and response collection were implemented in MATLAB (The MathWorks, Natick, MA, USA) using the Psychophysics Toolbox \cite<PTB--3;>{Brainard.SpatialVision.1997,Kleiner.Perception.2007,Pelli.SpatialVision.1997} and Eyelink Toolbox \cite{Cornelissen.BehavResMethInsC.2002}.

\subsection*{Stimulus material}
\begin{figure*}
\begin{center}
\includegraphics[width=\linewidth]{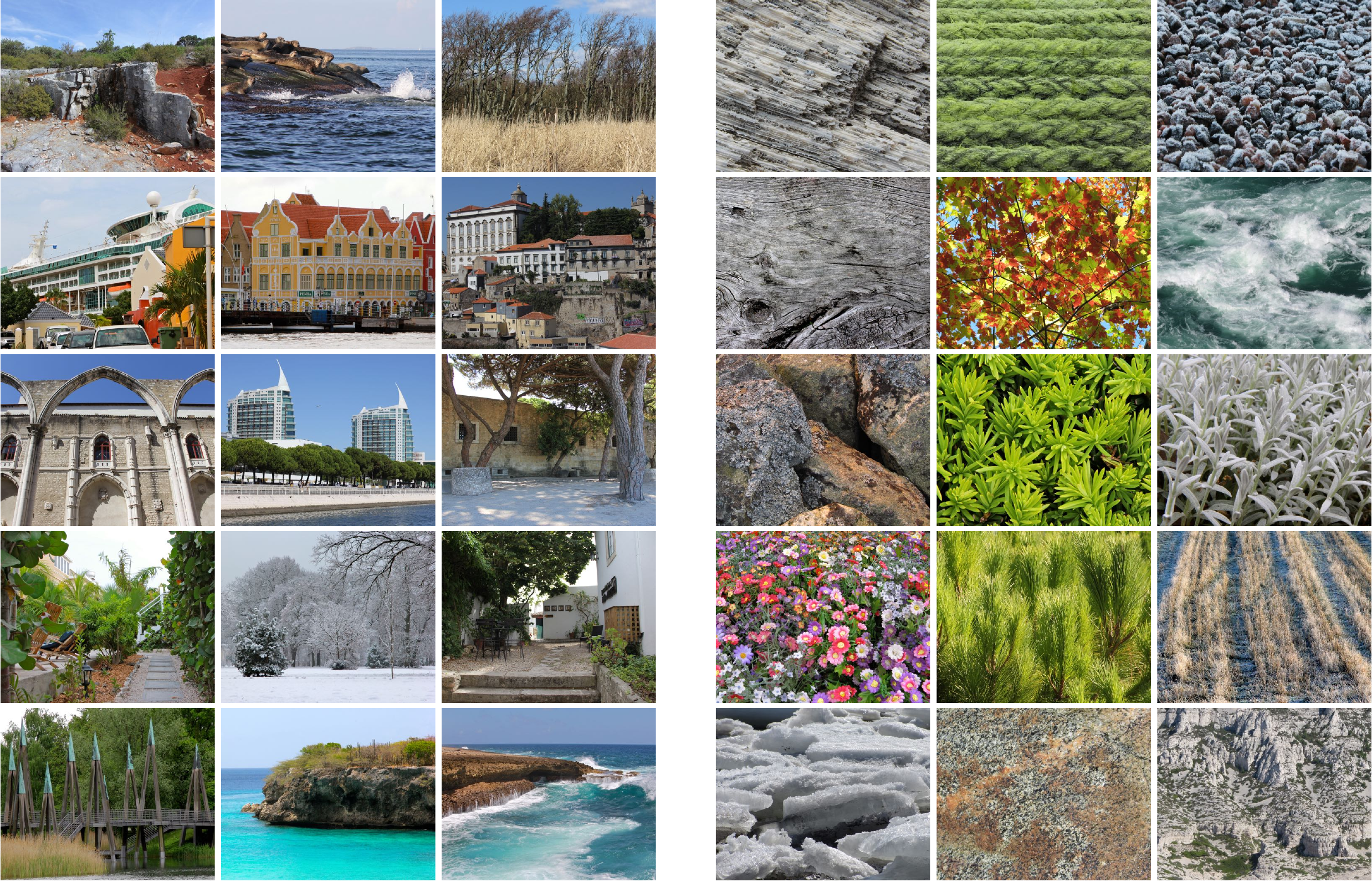}
\caption{Images used in experiment. Left: natural scenes; right: texture images.}
\label{images}
\end{center}
\end{figure*}

We used two sets of colored photographs in our experiment. In the first part of the experiment each set consisted of 15 images (Fig.~\ref{images}). Image Set 1 contained photographs of natural landscapes and rural scenes. Image Set 2 contained photographs of textures. During the memory test in the second part of the experiment we added 15 novel images of the same category to each image set.

\subsection*{Task and procedure}
Each trial began with the presentation of a black fixation cross in front of a uniform gray background. The fixation cross was placed randomly within the image boundaries. After successful fixation, the fixation cross was replaced by the image for 10~s.  Participants were instructed to explore each image for a subsequent memory test. The first block of the experiment consisted of 60 trials where each image was presented twice. Results from the first inspection of images have been described previously \cite{Engbert.JVis.2015}. The second inspection has not been published earlier. In a second block participants completed a memory test with 60 trials. The memory test contained all presented images and thirty new images with natural scenes and textures. Participants had to judge whether the  image had been presented during Block 1.

To minimize the potential influence of the monitor frame and since accuracy of eye trackers falls off towards the edges of a monitor, images were presented centrally with gray borders extending 32 pixels to the top/bottom and 40 pixels to the left/right of the image. The resulting size of the image was 1200$\times$960 pixels (31.1\degree$\times$24.9\degree). 

\subsection*{Data preprocessing}
We detected saccades by using a velocity-based algorithm \cite{Engbert.VisionRes.2003,Engbert.ProcNatlAcaSciUSA.2006}. Saccades were defined as fast movements of both eyes that exceeded the average velocity during a trial by 6 standard deviations for at least 6~ms with a minimum amplitude of 0.5$^\circ$. Eye traces between successive saccades were tagged as fixations. Fixation positions were computed by averaging the mean eye position of both eyes. Since trials started with a fixation check, first fixations on images were removed from the data set ($N=2100$). In addition, fixations containing a blink or with a blink during an adjacent saccade were excluded from subsequent analyses ($N=2214$). Overall, 55.526 fixations remained for further analyses.

\subsection{SceneWalk model}
For the interpretation of our results, we simulated fixation sequences with the SceneWalk model (\citeNP{Engbert.JVis.2015}; cf., \citeNP{Schutt.PsycholRev.2017}). Fixation sequences are generated in the computational model by two competing activation maps: (i) an excitatory attention map that provides potential saccade targets and (ii) an inhibitory fixation map that tags previously fixated locations. Maps in the model are discretized with dimensions $k \times l = 128 \times 128$ \cite{Stensola.Nature.2012}. Activations in the attention map $a_{ij}$ at coordinates $(i,j)$ evolve over time. To approximate potential saccade targets we compute the empirical density of fixation locations on a given image. The empirical density contains all effects generated by bottom-up and top-down processing that contribute to the inhomogeneous distribution of fixation locations. In our model simulations, the empirical density feeds into the attention map. Extraction from the empirical density is highest at fixation and decreases with increasing eccentricity. This corresponds to the attentional window of our model. Mathematically, the empirical density of fixation locations is weighted by a Gaussian envelope of size $\sigma_1$. Position of the attentional window $a_{ij}$ changes after each saccade and remains constant otherwise. In addition leakage leads to a temporal decay of activations. The updating rule of the attention map $a_{ij}$ is given by


\begin{equation}
a_{ij}(t+1) = \frac{\phi_{ij}A_{ij}(t)}{\sum\limits_{kl} \phi_{kl}A_{kl} (t)} + (1-\rho) a_{ij}(t)
\label{eqAttMap}
\end{equation}

\noindent with the 2D-Gaussian $A_{ij}(t)$ centered upon fixation at time $t$, the distribution of fixation locations $\phi_{ij}$, and the rate of decay $\rho$. 

Temporal evolution of activations in the inhibitory map $f_{ij}$ is very similar to the dynamics in the attention map $a_{ij}$. Temporal evolution consists of activation accumulation centered at fixation and a proportional temporal decay across the map. The updating rule for the fixation map $f_{ij}$ is given by
\begin{equation}
f_{ij}(t+1) = F_{ij}(t) + (1-\omega) f_{ij}(t)
\label{eqFixMap}
\end{equation}

\noindent with a 2D-Gaussian $F_{ij}(t)$ with standard deviation $\sigma_0$ centered at fixation at time $t$ and a decay rate of the fixation map $\omega$. The fixation map tracks fixated areas and is motivated by inhibition of return \cite{Klein.PsycholSci.1999}, which has been suggested as a mechanism to drive exploration in scenes \cite{Itti.NatRevNeurosci.2001}. Although the role of inhibition of return has been questioned \cite{Smith.VisCogn.2009}, model simulations support inhibitory tagging as an important mechansim during scene perception (\citeNP{Rothkegel.VisionRes.2016}; cf. \citeNP{Bays.JVis.2012}).

While attention and fixation map evolve independently over time, both maps are subsequently normalized and combined into a single map for target selection. The potential $u_{ij}$ is given by

\begin{equation}
u_{ij}(t) = - \frac{[a_{ij}(t)]^\lambda}{\sum\limits_{kl} [a_{kl}(t)]^\lambda} +
	\frac{[f_{ij}(t)]^\gamma}{\sum\limits_{kl} [f_{kl}(t)]^\gamma}
\label{eqPotential}
\end{equation}

\noindent where the exponents $\lambda$ and $\gamma$ are free parameters. \citeA{Engbert.JVis.2015} fixed these parameters to $\lambda = 1$ to reproduce the densities of gaze positions and $\gamma = 0.3$ to reproduce spatial correlations between fixation locations. We kept these values in our simulations. The probability of a location $(i,j)$ to be chosen as the next saccade target can be extracted from the potential, i.e.

\begin{equation}
\pi_{ij}(t) = \max \left( \frac{u_{ij}(t)}{\sum\limits_{(k,l) \in S } u_{kl}(t)} , \eta \right)
\label{eqTarSel}
\end{equation}

\noindent where $S$ contains all positions on the grid with $u_{ij}(t) \le 0$ and a free parameter $\eta$ that adds noise to the selection process so that every position has at least a minimal probability to be chosen as the next saccade target. Target selection in the SceneWalk model occurs at the end of fixation where the eyes move instantaneously. The intervals between successive saccades were drawn from a Gamma distribution with a shape parameter of $9$ and a scale parameter of $\sim\!0.031$, which corresponds to a mean fixation duration of 275~ms.

\subsection{Parameter estimation}
Simulations were based on different parameters for each of the four experimental conditions (presentation $\times$ image type). As a starting point we used the parameters reported in \citeA{Engbert.JVis.2015}. These parameters were estimated for fixations during the first presentation of natural images. We hypothesized that image type (natural scenes vs. textures) might affect target selection and decided to estimate all parameters for the first presentation of texture images anew. Previous work suggested that reinspection of images leads to a decreased attentional span \cite{Kaspar.PLoSOne.2011,Kaspar.JVis.2011}. Hence, we decided to fix all parameters except for the sizes of the attention span $\sigma_1$ and inhibition span $\sigma_0$ for simulation of the second inspection of images. We used a genetic algorithm approach \cite{Mitchell.BOOK.1998} to estimate model parameters. Parameter estimation was based on the first five images for each image type. The remaining ten images in each image set were used for model evaluations. Limiting the analysis to the predicted images did not alter effects. We used first-order statistics (2D density of fixation locations) and the distribution of saccade lengths as an objective function to evaluate parameters. A list of estimated parameter values and standard errors can be found in Table~\ref{SceneWalkPar}. 

\begin{table}
\caption{Model parameters. Standard errors were calculated from five parameter estimations of the experimental data recorded during the first presentation of natural images.}
\label{SceneWalkPar}
\begin{center}
\begin{tabular}{lcccccc}
\hline
& & \multicolumn{2}{c}{Natural Images} & \multicolumn{2}{c}{Texture Images} \\
& Parameter & First & Second & First & Second  & Error\\
\hline
Attention span & $\sigma_1$ & 4.88 & 4.20 & 4.72 & 4.20  &  $\pm$0.11\\
Attention decay & $\log_{10}\rho$  &  -1.18 &  -1.18 & -1.36 & -1.36 & $\pm$0.28 \\
Inhibition span & $\sigma_0$  & 2.16 & 2.18 & 1.79 & 1.76 & $\pm$0.25\\
Inhibition decay & $\log_{10}\omega$ & -4.03 & -4.03 & -3.80  &  -3.80 & $\pm$0.08 \\ 
Noise & $\log_{10}\eta$ & -4.04 & -4.04 & -4.51 &  -4.51 & $\pm$0.07 \\
\hline
\end{tabular}
\end{center}
\end{table}

\htr{
\subsection{Control Model: Joint Probability of Saccade Amplitude and Fixation Density}
To investigate the influence of saccade amplitudes on the aggregation of points, we simulated a control model that generates both realistic distributions of fixation positions  and saccade amplitudes. Fixation sequences began at the initial fixation position of the empirically observed scanpath and subsequent fixation positions were simulated iteratively. The next fixation position was chosen proportional to the joint probability of the empirical density of saccade amplitudes and the empirical density of fixation positions on an image. For each empirical fixation sequence we simulated a scanpath with the same number of fixations to exclude effects due differences in the number of fixations or sequences. The optimal bandwidth for density estimation was chosen by applying Scott's rule of thumb (\texttt{bw.scott}) for fixation densities and unbiased cross-validation (\texttt{bw.ucv}) for saccade amplitudes.

\subsection{Statistical modeling}
For statistical analyses, we computed linear mixed effect models for each dependent variable using the \emph{lme4} package \cite{Bates.JStatSoft.2015} in \emph{R} \cite{RCoreTeam.2018}. We \emph{log}-transformed both dependent variables, since they deviated considerably from normal distributions. For the statistical model of the empirical data, we used the maximal possible random effect structure \cite{Barr.JMemLang.2013} and ensured that none of the models was degenerate \cite{Bates.arXiv.2015}. For our results we interpret all $|t|$$>$2 as significant fixed effects \cite{Baayen.JMemLang.2008}. 

In R the linear mixed effect model for the empirical data can be written as

\begin{equation}
\mbox{\texttt{log(dv) $\sim$ }} 
\underbrace{\mbox{\texttt{1 + A + B + A:B }}}_{\mbox{fixed effects}}
\underbrace{\mbox{\texttt{+ ( 1 + A + B + A:B | id )}}}_{\mbox{random effects id}}
\underbrace{\mbox{\texttt{+ ( 1 + A | img )}}}_{\mbox{random effects image}}
\label{eq_lme}
\end{equation}
\noindent with a fixed effects of presentation \texttt{A}, image type \texttt{B} and their interaction \texttt{A:B} and the corresponding random effect structure. As each image belongs to only one image type, image type is a between factor for images and was not included in the random effect structure for images.

The random effect structure for the simulated data differed from Equation~\ref{eq_lme}, since fixation sequences were only based on one participant, that is, the average participant. Therefore, we did not estimate random effects for participants. Due to convergence problems we also removed the random slope of presentation for each image. This reduced the random effect structure to a single intercept per image for the simulated data. 
}

\section*{Results}
With our first analysis we expected to replicate the results of \citeA{Kaspar.PLoSOne.2011,Kaspar.JVis.2011} that a second inspection of the same image leads to decreased saccade amplitudes. In addition, we investigated whether the two image types (natural scenes vs.~textures)  differentially affected saccade amplitudes. In a second analysis, we tested the sensitivity of the \emph{pair correlation function} (PCF) to our experimental manipulations. All experimental results were compared to model simulations of the SceneWalk model \cite{Engbert.JVis.2015} \htr{and the joint probability control model that reproduces fixation densities and saccade amplitudes. In addition, we display the results of the inhomogeneous and the homogeneous point processes. These processes are plotted to demonstrate that results of the PCF are not generated by the method itself.}

\subsection*{Saccade amplitudes}
\begin{figure*}[htb]
\begin{center}
\includegraphics[width=.9\linewidth]{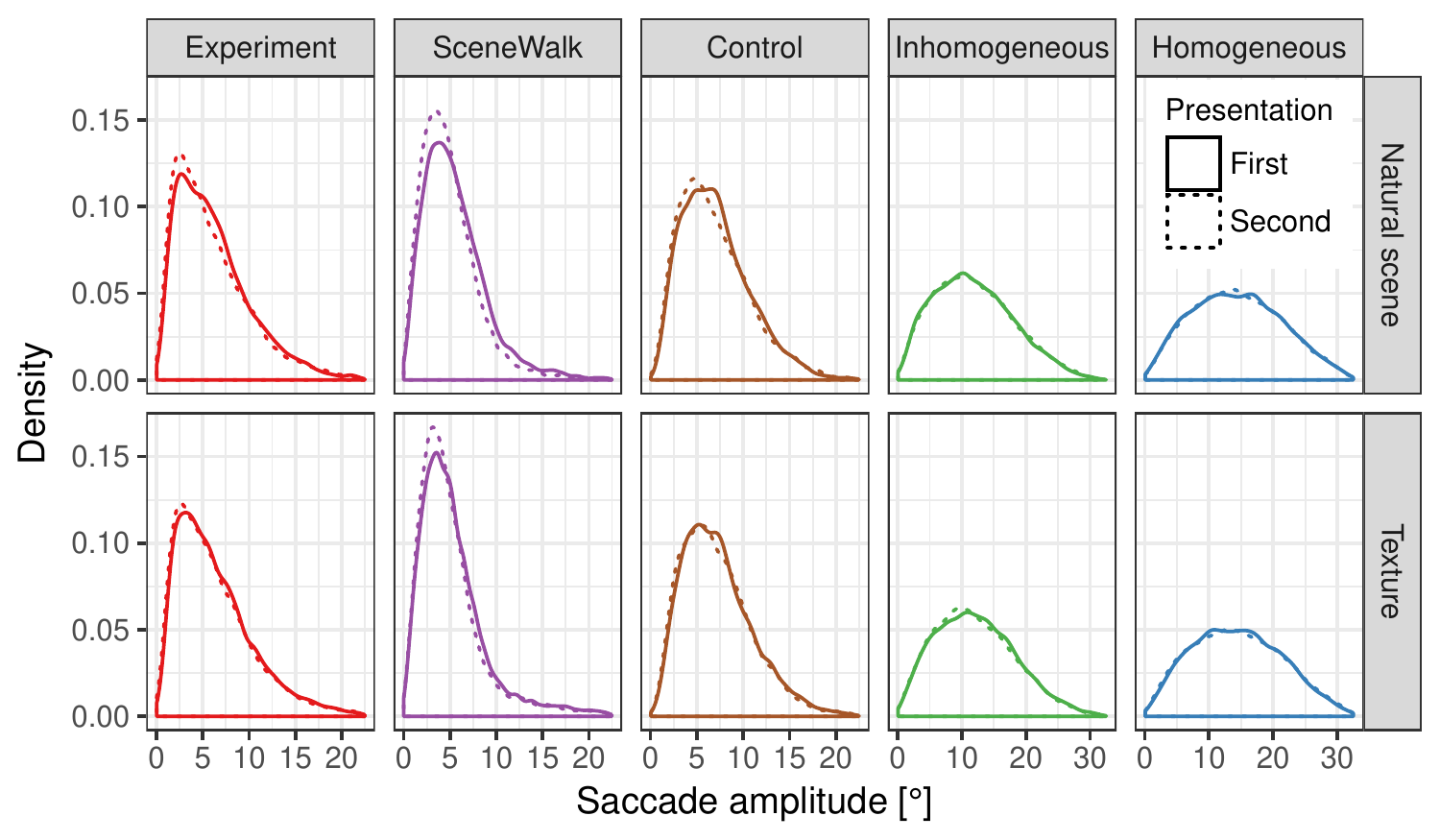}
\caption{\htr{Saccade amplitudes. Experiment (red), and results of model simulations by the SceneWalk model (purple), a joint probability model of saccade amplitudes and fixation density (brown), an inhomogeneous point process (green), and an homogeneous point process (blue). Note, scales of the x-axis differ between the first three and the last two point processes.}}
\label{SacAmpDist}
\end{center}
\end{figure*}

Figure~\ref{SacAmpDist} shows \htr{the distribution of  saccade amplitudes} on natural scenes (top) and texture images (bottom) during the first and second presentation \htr{(solid vs. dashed lines). As expected, empirical saccade amplitudes distributions are positively skewed and long-tailed. Distributions of saccade amplitudes of the SceneWalk model and the joint probability control model deviate less from the empirically observed saccade amplitudes than the inhomogeneous and homogeneous point processes. At closer inspection, the SceneWalk model generates slightly shorter amplitudes while the joint probability control model generates slightly longer saccade amplitudes than our participants. 

A linear mixed effects model (LME) for the experimental data \cite<see Methods section; >{Bates.JStatSoft.2015} revealed a significant fixed effect of Presentation but no effect of Image Type and no interaction (Tab.~\ref{tabLmeSacAmp}, left columns). Saccades were larger during the first inspection than during the second inspection. LME models of saccade amplitudes generated by the SceneWalk model and the joint probability control model replicated this effect qualitatively. However, the interaction of Presentation and Image Type  also reached significance for the SceneWalk model. The reduction of saccade amplitudes during the second inspection was stronger on natural scenes than on texture images. 
}

\begin{table}
\caption{Fixed effects of linear mixed effect models. 
      For each point process we estimated separate models for saccade amplitudes (\emph{log}-transformed) and 
      the summed deviation from complete spatial randomness of the PCF (\emph{log}-transformed). 
      The table reports estimates of fixed effects 
      ($\beta$) with standard errors (SE) and $t$ values. $|t|>2$ are interpreted as significant effects.}
\label{tabLmeSacAmp}
\begin{center}
\begin{tabular}{lcccccccc}
\hline
& \hspace{.1cm} & \multicolumn{3}{c}{Saccade amplitudes} & \hspace{.3cm} &  \multicolumn{3}{c}{PCF deviation} \\ 
& & $\beta$ & SE & $t$ & & $\beta$ & SE & $t$\\
\hline
\bf{Experiment}\\
(Intercept) & & 1.599 &  0.034 & 46.58 & &  5.668 &  0.095 & 59.43 \\
Presentation & & -0.062 &  0.012 & -5.33 & &  0.508 &  0.099 & 5.13 \\
Image Type & & 0.035 &  0.023 & 1.52 & &  -0.073 &  0.123 & -0.59 \\
Presentation $\times$ Image Type & & 0.024 &  0.016 & 1.51 & &  -0.144 &  0.187 & -0.77 \\\\
\bf{SceneWalk}\\
(Intercept) & & 1.459 &  0.006 & 232.72 & &  5.519 &  0.037 & 147.80 \\
Presentation & & -0.090 &  0.006 & -16.18 & &  0.586 &  0.062 & 9.52 \\
Image Type & & -0.010 &  0.013 & -0.83 & &  -0.135 &  0.075 & -1.81 \\
Presentation $\times$ Image Type & & 0.058 &  0.011 & 5.25 & &  -0.211 &  0.123 & -1.71 \\\\
\bf{Joint Probability Control}\\
(Intercept) & & 1.805 &  0.011 & 165.32 & &  5.123 &  0.030 & 168.32 \\
Presentation & & -0.033 &  0.005 & -6.39 & &  0.219 &  0.055 & 4.00 \\
Image Type & & 0.040 &  0.022 & 1.81 & &  0.168 &  0.061 & 2.75 \\
Presentation $\times$ Image Type & & 0.010 &  0.010 & 0.94 & &  -0.114 &  0.110 & -1.04 \\
\hline
\end{tabular}
\end{center}
\end{table}

\subsection*{Second order statistics: Pair correlation function}
\begin{figure*}
\begin{center}
\includegraphics[width=.9\linewidth]{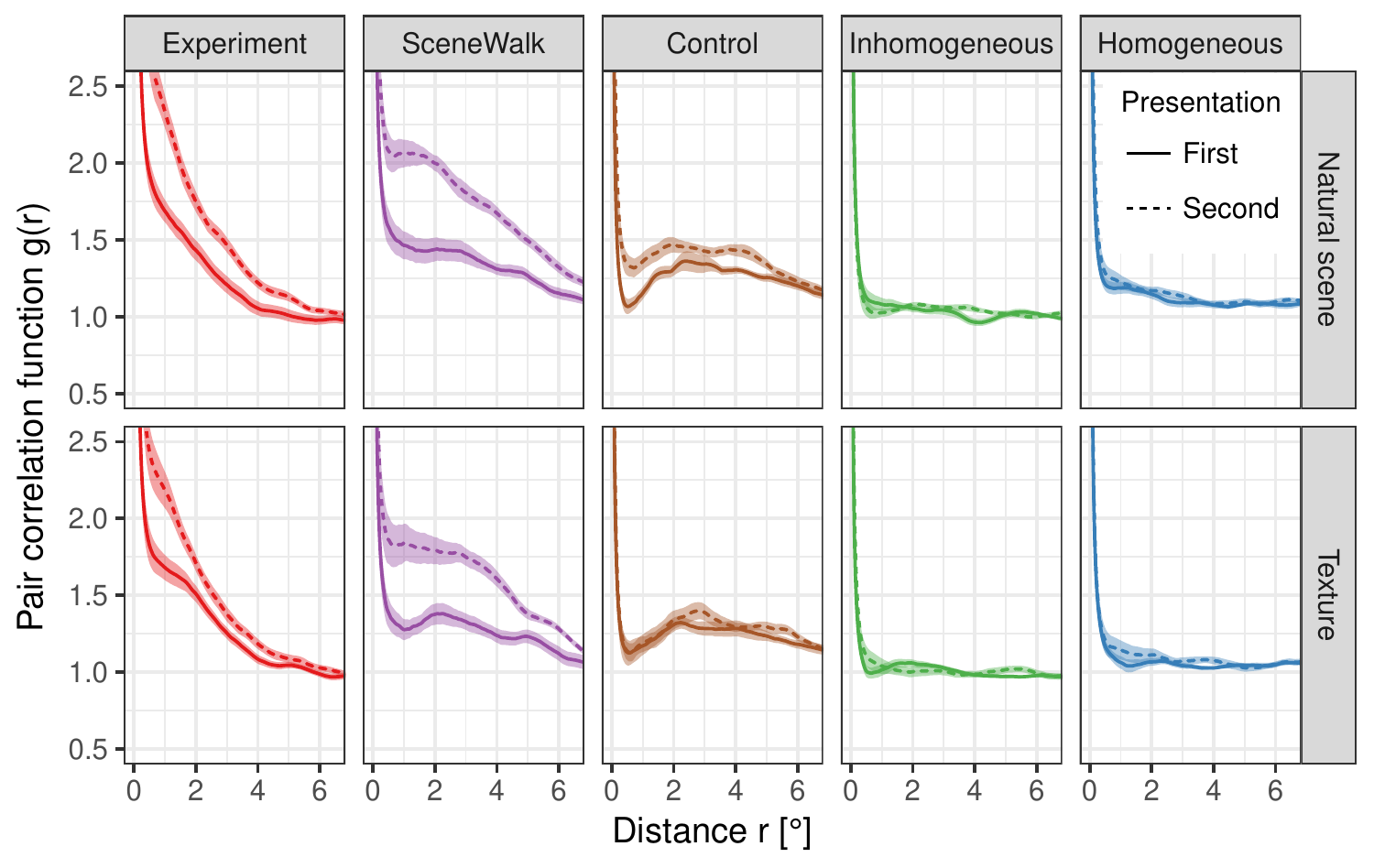}
\caption{Average pair correlation functions (PCFs) for experimental and simulated point processes. PCFs were estimated for each condition separately (natural vs. texture images: top vs. bottom row;  first vs. second inspection: solid vs. dashed lines). Estimated PCFs for the observed fixation locations (red lines), and simulated locations generated by the SceneWalk model (purple lines), a joint probability model of saccade amplitudes and fixation density (brown lines), an inhomogeneous point process (green lines), and an homogeneous point process (blue lines). 95\% confidence intervals represent the variability across images.}
\label{pcfAll}
\end{center}
\end{figure*}

\begin{figure*}
\begin{center}
\includegraphics[width=.9\linewidth]{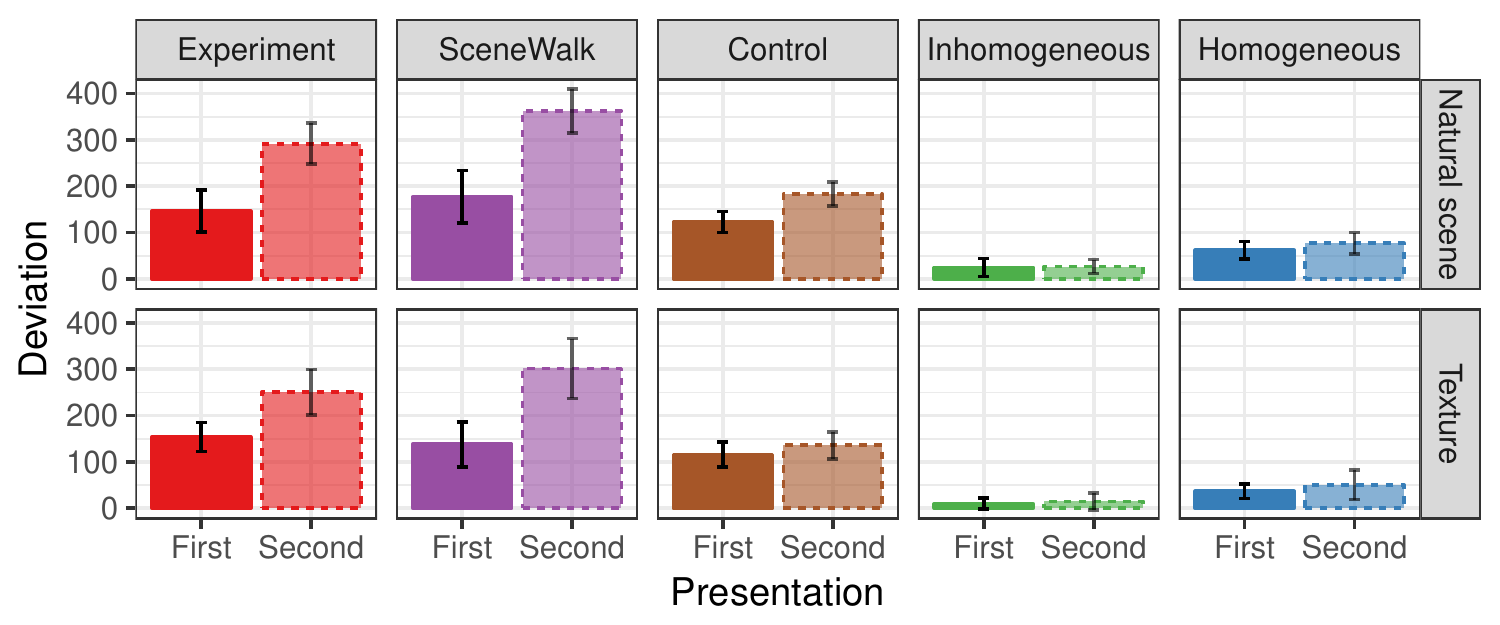}
\caption{Summed deviation from complete spatial randomness (Eq.~\ref{eqTestStat}) of empirical and simulated point processes for all conditions (natural vs.~texture images;  first vs.~second inspection). The deviation was computed for distances $0.1\degree<r<6.5\degree$. As expected, the deviation from complete spatial randomness is smallest for the two control processes (homogeneous and inhomogeneous point process).}
\label{pcfAllDev}
\end{center}
\end{figure*}

\htr{We computed inhomogeneous pair correlation functions (PCFs) for each condition of the five point processes (Fig.~\ref{pcfAll}).} We observed spatial correlations of fixation locations during individual trials in all conditions of our experimental data (red lines). Fixations locations were more abundant than expected from the overall inhomogeneity of fixation locations at short distances~$r$. The estimated PCFs deviated from complete spatial randomness, i.e., $ \hat{g}(r) > 1$, at distances $r < 4\degree$. More importantly, the second presentation of an image (dashed lines) led to increased PCFs for both natural scenes (top row) and texture images (bottom row). Statistically, we evaluated spatial correlations by computing the deviation from complete spatial randomness of each PCF, i.e., the summed deviation of the PCF from one for distances $0.1 \le r \le 6.5$ (Fig.~\ref{pcfAllDev}; cf.~Eq.~\ref{eqTestStat}). A linear mixed effect model (LME) revealed a significant deviation from complete spatial randomness (intercept) and an effect of  Presentation (Tab.~\ref{tabLmeSacAmp}). \htr{All other fixed effects were non-significant.} Thus, deviations \htr{from complete spatial randomness} were present in all conditions of our experiment with larger deviations during the second inspection of an image \htr{irrespective of image type.}

The SceneWalk model replicated this pattern of results qualitatively (Fig.~\ref{pcfAll}, purple lines). All conditions showed strong spatial correlations. PCFs deviated from \htr{complete spatial randomness} for distances $r < 6\degree$. The effect extended to larger distances in our model simulations \htr{than in the experimental data.} We analyzed deviations from \htr{complete spatial randomness} with another LME for the SceneWalk model (Tab.~\ref{tabLmeSacAmp}; cf.~Fig.~\ref{pcfAllDev}). PCFs deviated from \htr{complete spatial randomness} (intercept). The effect was larger for the second inspection. \htr{No other fixed effect was significant for the deviation score of the SceneWalk model.}

\htr{In order to check whether spatial correlations between fixation locations are primarily generated by the tendency of participants to generate short saccade amplitudes, we simulated a control model based on the joint probability of saccade amplitudes and the distribution of fixations (see Methods). The PCF of the control model showed correlations at all evaluated distances. In direct comparison to the experimental data the control model generated considerably weaker correlations at small distances $r <3\degree$ and generated stronger correlations at large distances $r > 4\degree$. Thus, a model that is solely based on the generation of realistic saccade amplitudes and fixation densities generates a qualitatively different correlation pattern. We analyzed deviations from complete spatial randomness with another LME for the control model (Tab.~\ref{tabLmeSacAmp}; cf.~Fig.~\ref{pcfAllDev}). PCFs deviated from complete spatial randomness (intercept). The effect was larger for the second inspection and on texture image. We observed no interaction of Presentation and Image Type in the deviation score.}
 
\htr{Finally, the average PCFs of the inhomogeneous point process indicated only small spatial correlations, i.e., $ \hat{g}(r) \approx 1$ at all distances  (Fig.~\ref{pcfAll}, green lines). The absence of spatial correlations for the average PCF was expected, since the optimal bandwidth $\lambda$ for the estimation of the PCF was chosen to minimize the deviation from complete spatial randomness of the inhomogeneous process (see Step 2 of the estimation process). PCFs of the homogeneous point process were similar to PCFs of the inhomogeneous point process, i.e., $ \hat{g}(r) \approx 1$ at all distances (Fig.~\ref{pcfAll}, blue lines).} 

\section{Discussion}
During scene perception, fixations are not \htr{uniformly} distributed on an image. Instead fixations cluster in parts of an image due to bottom-up factors, top-down factors, and systematic tendencies of gaze control \cite{Tatler.JEMR.2008}. We propose to use the \emph{pair correlation function} (PCF) to investigate the relation of fixation positions within single trials and demonstrate \htr{that the PCF reveals aggregation of points at distances $r<4\degree$, that is, it is more likely to observe fixation locations in the proximity of another fixation location than expected by the overall distribution of fixation locations. This effect cannot be explained by the tendency of participants to generate short saccade amplitudes alone, as simulations of a control model that samples fixation locations from the joint probability of the density of saccade amplitudes and the density of fixation locations led to a qualitatively different PCFs. The control model underestimated aggregation at short distances and overestimated aggregation at large distances. In addition, the PCF responded sensitively to a memory manipulation in our experiment and revealed stronger aggregation of points during the second inspection than during the first inspection of an image.} Simulations of the SceneWalk model \cite{Engbert.JVis.2015} demonstrated that a reduced attentional span \htr{could lead to reduced saccade amplitudes and} explain the observed results of the PCF \htr{during the second inspection.}

\subsection{Pair correlation function}
Research on eye movements during scene perception has focused on the distribution of fixation locations across observers, as for example during the evaluation of saliency models \cite{Bylinskii.SaliencyBenchmark}. However, this approach neglects dependencies between fixation locations during a trial \cite{Engbert.JVis.2015}. The PCF is a method from spatial statistics to evaluate the relation of pairs of points \cite{Diggle.Book.2013} and reveals whether points solely depend on the inhomogeneity of a point process or whether pairs of points affect each other mutually at a given distance $r$. The PCF can be applied to eye movement data (i.e., fixation locations) in three steps. In a first step, inhomogeneous and homogeneous point processes need to be simulated to evaluate the PCF estimation. Both point processes generate fixation locations that are independent at all distances~$r$. Hence, PCFs of both processes are expected to show no spatial correlations at any distance~$r$. During a second step an optimal bandwidth needs to be chosen for the estimation of the PCF. As a criterion we suggest to use a bandwidth for which the PCF of the \htr{simulated} inhomogeneous point process has the least deviation from complete spatial randomness, i.e.  shows no spatial correlations. In a last step, the computed bandwidth is used to compute PCFs for each individual trial. 

In scene perception, the PCF characterizes whether fixation locations can be explained by the underlying distribution of all fixation locations (no spatial correlations between pairs of points) or whether fixation positions interact with each other at distance $r$. PCFs revealed spatial correlations of fixation locations in all conditions in our experimental data. Fixations were more abundant at distances $r<4\degree$ than we would expect from the inhomogeneity of fixation locations alone. Beyond $4\degree$ fixation locations were independent of each other. Thus, observing a fixation increased the probability of observing more fixations than expected by the overall inhomogeneity within $4\degree$. Beyond $4\degree$ fixations were as likely as predicted by the local intensity of fixation locations. As expected, neither the inhomogeneous nor the homogeneous point process revealed strong spatial correlations, since fixation locations were independent of each other for these point processes. Therefore, aggregation observed in our PCFs is generated \htr{by the empirical data and is not the result of the method itself.}
 
\htr{Finally, the PCF provides a quantitative statistic of spatial correlations and can be used to compare data sets generated by different processes or under different conditions. The data sets may even differ in the overall inhomogeneity of the fixation location distribution, since the inhomogeneous PCF takes this inhomogeneity into account. Hence, we were able to compare point patterns of the first and second inspection of an image, point patterns on different image types as well as empirical and simulated point patterns. Even though the distribution of fixation locations and the distribution of saccade amplitudes were similar among these data sets, the PCF revealed considerable differences in the correlation patterns. For example, the experimental data, simulations of the SceneWalk model and simulations of the joint probability control model unveiled strong differences in spatial correlations. While the SceneWalk model generated stronger but qualitatively similar spatial correlations when compared to the experimental data, the joint probability control model produced a different pattern. Fixation locations aggregated at all evaluated distances, but produced much weaker correlations for short distances. Thus, short saccade amplitudes are not sufficient to produce the strong correlations at distances $r<2\degree$. There must be additional mechanisms that let participants fixate the same locations in a scanpath as through direct regressions \cite<cf. facilitation of return; >{Smith.VisCogn.2009} and through reinspections later during a trial. 

Interestingly, image type did not affect spatial correlations. However, we only tested two types of images and a broader range of image types is needed to see the generalizability of this result. To conclude the PCF is a powerful tool to compare spatial correlations of point patterns and will help to understand the eye-movement dynamics under different experimental conditions as well as the dynamics of different models of eye-movement control during scene perception.}

\subsection{Repeated presentation of images}
To test the sensitivity of the PCF, we recorded eye movements of participants while viewing an image twice. The repeated presentation was expected to result in shorter saccade amplitudes due to a reduced attentional span \cite{Kaspar.PLoSOne.2011,Kaspar.JVis.2011}. We replicated shorter saccade amplitudes during the second inspection independent of image type \htr{in the experimental data.} In addition, we observed stronger spatial correlations within $4\degree$ during the second inspection. 

Reduced saccade amplitudes as well as increased aggregation might be generated by a reduced attentional window \cite{Kaspar.PLoSOne.2011,Kaspar.JVis.2011}. We tested this hypothesis with simulations of the SceneWalk model \cite{Engbert.JVis.2015}. Parameter estimation led to a reduced attentional span in the SceneWalk model  during the second inspection of images which in turn led to shorter saccade amplitudes and stronger aggregation of fixation locations as quantified by the PCFs. Hence, our simulation results are in agreement with the interpretation of a reduced attentional span during repeated inspection of images. 

In its current form the SceneWalk model overestimates aggregation in particular during the second inspection. However, our results are predictions and are not optimized to account for the experimentally observed PCFs. Beside the size of the aggregation, we observed a difference in the functional form of the PCF. PCFs of the SceneWalk model did not decrease as fast as those estimated from experimental data. This might have resulted from the Gaussian form of our attentional window with its long tails. A revision of the model using a modified attentional window might improve model fit in this respect \cite<cf.,>{Schutt.PsycholRev.2017}.


\htr{
\subsection{Relation to other metrics}
Several measures can be used to study the dynamics of scanpaths. The measure that is most strongly related to the PCF is based on Voronoi diagrams \cite{Over.BehavResMeth.2006}. The suggested method provides a measure for the uniformity of a pattern of fixation locations and is normalized by the number of fixation locations. Hence, the measure can be used to compare the uniformity of different data sets. While the Voronoi method allows to estimate the inhomogeneity of a point pattern, the inhomogeneous PCF takes this inhomogeneity into account and reveals spatial correlations between points that cannot be explained by the inhomogeneity. Most importantly, the inhomogeneous PCF can be used to compare point patterns with different levels of inhomogeneity. Thus, the Voronoi method and the PCF complement each other.  

A number of scanpath comparison methods have been proposed that provide metrics to describe the similarity of the dynamics of two scanpaths \cite<for a review see>{Anderson.BehavResMeth.2015}. Each metric quantifies unique aspects of the scanpaths and helps to understand which aspects  resemble each other or differ between two scanpaths. However, the focus of all methods is the comparison of scanpaths and requires matching scanpaths. In contrast, the PCF does not require pairs of scanpaths as each scanpath is evaluated by its own. This allows to compare arbitrary scanpaths. In addition, PCF and scanpath comparison metrics might come to very different conclusions. Since the PCF describes spatial correlation of points, two completely different point patterns might produce the same correlations. At the same time, two seemingly similar point patterns might result in very different spatial correlations.

Other approaches have described a large variety of individual effects that influence eye movement behavior during scene perception \cite<e.g.,>{Smith.VisCogn.2009,Tatler.JVis.2007,Tatler.JEMR.2008} and best practices have been suggested to estimate appropriate baselines for some of them, as for example for the central fixation bias \cite{Clarke.VisionRes.2014,Clarke.JVis.2017}. Each individual effect describes parts of the dynamics present in eye movements, but none is a sufficient metric to capture the overall complexity. Since the PCF takes all fixations of a sequence into account, it uncovers the consequences of these effects that remain unnoticed on shorter time scales. Hence, the PCF acts in concert with these other methods to understand eye movements during scene perception.
}

\subsection*{Conclusions}
The \emph{pair correlation function} (PCF) is a powerful tool to analyze spatial correlations of fixation locations. During scene perception the PCF reveals aggregation of fixations during individual trials and reacts differentially to experimental manipulations. Simulations of a computational model demonstrate that a reduced attentional span leads to increased aggregation of fixation locations. Our work provides an example how spatial statistics and computational modeling can be combined to investigate general statistical properties of eye movement control.

\subsection*{Acknowledgments}
This work was funded by Deutsche Forschungsgemeinschaft (DFG, German Research Foundation, grant EN 471/13--1 to R.E.) and the Collaborative Research Centre SFB 1294, project number B05 and by Bundesministerium f\"ur Bildung und Forschung (BMBF) via Bernstein Center for Computational Neuroscience Berlin (Project B3, F\"orderkennzeichen 01GQ1001F to R.E. \& F.A.W.). Simon Barthelme was supported by an ANR grant (GenGP, ANR-16-CE23-0008).

%
%
\bibliographystyle{apacite}
\bibliography{lib.bib}

\appendix

%
%

\end{document}